# Measurement Accuracy in Silicon Photonic Ring Resonator Thermometers: Identifying and Mitigating Intrinsic Impairments

**S. Janz,[1] S. Dedyulin,[2] D.X. Xu,[1] M. Vachon,[1] S. Wang,[1] R. Cheriton,[1] and J. Weber[1]**

[1]*Advanced Photonics and Electronics Research Centre, National Research Council Canada, Ottawa, Canada, K1A 0R6*
[2]*Metrology Research Centre, National Research Council Canada, Ottawa, Canada, K1A 0R6*
*\*Siegfried.janz@nrc-cnrc.gc.ca*

**Abstract:** Silicon photonic ring resonator thermometers have been shown to provide temperature measurements with a 10 mK accuracy. In this work we identify and quantify the intrinsic on-chip impairments that may limit further improvement in temperature measurement accuracy. The impairments arise from optically induced changes in the waveguide effective index, and from back-reflections and scattering at defects and interfaces inside the ring cavity and along the path between light source and detector. These impairments are characterized for 220 × 500 nm Si waveguide rings by experimental measurement in a calibrated temperature bath and by phenomenological models of ring response. At different optical power levels both positive and negative light induced resonance shifts are observed. For a ring with L = 100 μm cavity length, the self-heating induced resonance red shift can alter the temperature reading by 200 mK at 1 mW incident power, while a small blue shift is observed below 100 μW. The effect of self-heating is shown to be effectively suppressed by choosing longer ring cavities. Scattering and back-reflections often produce split and distorted resonance line shapes. Although these distortions can vary with resonance order, they are almost completely invariant with temperature for a given resonance and do not lead to measurement errors in themselves. The effect of line shape distortions can largely be mitigated by tracking only selected resonance orders with negligible shape distortion, and by measuring the resonance minimum wavelength directly, rather than attempting to fit the entire resonance line shape. The results demonstrate the temperature error due to these impairments can be limited to below the 3 mK level through appropriate design choices and measurement procedures.

## 1. Introduction

A silicon photonic thermometer can be based on a Si waveguide ring resonator (RR) or photonic crystal resonator, where the temperature T is determined from the change in the resonance wavelengths [1-6]. The relatively large thermo-optic coefficient of the silicon waveguide core ($\gamma_{TO} \sim 2 \times 10^{-4}$) [7] results in a resonance wavelength variation with temperature $d\lambda_{res}/dT = (\lambda/N_g)(dN_{eff}/dT)$, where $N_g$ is the waveguide group index and $N_{eff}$ is the effective index. In silicon waveguides this temperature dependence is approximately 80 pm/K, with the exact value depending on the specific device geometry and operating wavelength. Wavelength changes down to 1 pm or less can be routinely measured using calibrated tunable lasers used in optical communication component testing, so a temperature resolution of better than 10 mK can be expected. Several national metrology organizations are assessing the performance of silicon photonic ring resonators for use in calibration laboratories and in high-accuracy commercial applications [4-6] as an alternative to platinum resistance thermometers (PRT), thermocouples, as well as a more precise alternative to fiber Bragg grating optical thermometers [1,2,4]. The interest in Si photonics arises in part because the waveguide is made using extremely pure single crystal silicon layer and encapsulated by a stable silicon dioxide cladding. As a result, a silicon photonic thermometer is expected to be both mechanically stable, shock resistant, and relatively immune to calibration changes caused by chemical contamination or changes of the constituent material properties over time and temperature cycling. Furthermore, Si RR thermometers are usually a few hundred micrometers or less in diameter, and therefore closely approximate a true point-like sensor, unlike PRTs and FBGs which are typically several centimeters long. This removes measurement ambiguities caused by thermal gradients across a large thermometer sensing element. Previously we have demonstrated a temperature measurement reproducibility of better than 10 mK is possible with our RR thermometer prototype using commercial optical test instruments [8]. In comparison, standard PRTs designed for demanding metrology applications are able to achieve reproducibility down to the 100 µK level [9]. This level of accuracy is a result of many decades of work on resistance measurement instrumentation and thermometer assembly, and best temperature-measurement practices. The result is the modern PRT with the Pt wire hermetically sealed with an appropriate gas mixture to suppress oxidation and surface contamination [9]. Achieving sub-mK accuracy and reproducibility with a Si photonic thermometer will require a similar assessment of measurement instrumentation, assembly, and intrinsic device properties, and development of strategies to limit calibration drift and measurement errors. A temperature resolution of 100 µK has already been demonstrated in a ring resonator by tracking a point at the side of a resonance using a constant power feedback loop to tune the probe laser wavelength [10], rather than measuring the resonance wavelength itself. This work demonstrates the potential of Si photonic thermometers, although the accuracy and repeatability of this interrogation method may not be suitable for a calibrated reference thermometer. In recent work, we have addressed some of the device assembly issues, for example by using a stress-free mounting for the Si thermometer chip and free space optical coupling to minimize stresses and contamination arising from adhesives bonding the Si chip to a substrate and to input/output optical fibers [4,8]. Using these approaches, photonic thermometer probe packages suitable for immersion in a temperature calibration bath have been built and are used here for the precise assessment of Si thermometer performance.

In this paper, we examine two basic physical impairment mechanisms that compromise Si photonic thermometer accuracy and reproducibility. The first arises from optically induced refractive index changes in the waveguide. Even if the incident optical power is very low, the circulating power in the ring on resonance can be orders of magnitude larger than the power in the through waveguide and produce measurable self-heating in the ring [2,11]. This is the analogous problem to that of resistive heating encountered in standard PRT thermometers [9]. Our results also reveal that at the low optical powers used to monitor the ring device, there are additional refractive index changes that are related to the dynamics of carrier excitation, recombination and trapping [11]. The second category of measurement impairments are caused by scattering and reflections at optical interfaces and defects within the chip and in the external optical path. The scattering and reflections can cause strong distortions in the ring resonance line shape, and create spurious fluctuations and ripple in the background spectrum of light transmitted through the thermometer chip which also distorts the apparent resonance line shape. These spectral distortions produce ambiguities in the measured resonance wavelength and hence temperature. Through experiment and modelling we quantify the potential measurement error that can result, and identify strategies for mitigation. These effects also contribute to detection limits and accuracy in other photonic sensor applications beyond thermometry, and so the results reported here should be of relevance to other types of silicon photonic sensors.

The paper is organized as follows. Section 2.1 gives an overview of the thermometer chip design and the thermometer probe assembly that is used for the temperature calibration bath measurements. The optical measurement and temperature measurement methods are described in section 2.2. A phenomenological model used to calculate resonator response to both temperature and incident power is summarized in Section 3. In Section 4 experiments and the phenomenological model are used to characterize and quantify the effect of optical self heating and other power dependent refractive index changes on ring performance. Section 5.1 examines the thermometer impairment arising from side-wall scattering and back-reflections within the ring cavity leading to line splitting and distortion and Section 5.1 addresses the effect of scattering and reflections elsewhere in the optical path. Section 6 summarizes the results and reviews the strategies and prospects for mitigating the various impairments examined in the previous sections.

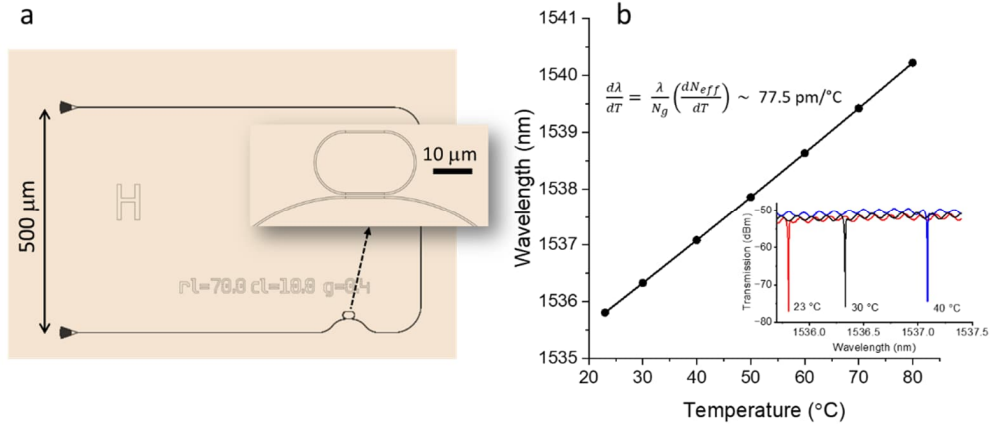

**Fig. 1.** (a) The layout of the ring resonator thermometer for Ring C. Input and output grating couplers are at left. The ring cavity length is L=70 μm. (b) The measured resonance wavelength variation for Ring C with temperature. The inset shows the resonance line at temperatures T = 23, 30, and 40 °C.

## 2. Experiment

### 2.1 Thermometer Chip Design and Probe Assembly

The ring resonators used in this work were formed using Si channel waveguides fabricated on silicon-on-insulator (SOI) wafers with a 220 nm thick Si waveguide layer and a 2 μm buried $SiO_2$ layer as the lower waveguide cladding. The rings and other waveguide structures were designed to guide TE polarized light at wavelengths near $\lambda$ = 1550 nm. The waveguides were formed by etching completely through the 220 nm Si layer to form a 500 nm wide waveguide. The upper waveguide cladding was a 2.2 μm thick PECVD deposited $SiO_2$ layer. In this work, measurement and theoretical modelling were carried out for four ring thermometers with different Si ring cavity lengths: Ring A with L=100 μm, Ring B with L =950 μm, Ring C with L = 70 μm, and Ring D with L =950 μm. Fig. 1 shows the ring resonator thermometer layout for Ring C, along with the measured temperature variation of resonance center wavelength and examples of the resonance spectrum taken at different temperatures. The other rings were similar in layout and design. The ring resonators were coupled to through-waveguides by directional couplers (DC) formed by two bends and a short parallel waveguide section as shown in Fig. 1. In the DC the two waveguides are separated by a nominal gap of 300 nm for Ring A, B and D, and 400 nm for Ring C, with parallel straight waveguide section lengths of 0 μm (Ring A), 5 μm (Ring B and Ring D), and 10 μm (Ring C). In the all-pass configuration shown in Fig. 1, the single through waveguide provides both optical input and output. The ring resonances appear as dips in the through waveguide transmission spectrum as shown in the inset of Fig. 1(b). Table 1 lists the selected resonance wavelength used in the thermometry experiments and corresponding ring properties for each device. Since the DC coupling coefficients and waveguide loss are sensitive to fabrication variations, the specific values given in Table 1 were extracted from experimental measurements. Light was coupled in and out of the through waveguide to/from two optical fibers by two focussing surface grating couplers at the left side of the chip in Fig. 1(a) [12].

Rings A and B were tested by placing the Si chips on a temperature-controlled optical test stage that is open to the laboratory environment. Rings C and D were each incorporated into a sealed probe with a schematic layout as shown in Fig. 2 [4, 8]. These probes were designed to be immersed in a temperature calibration bath at the Metrology Research Centre at the National Research Council Canada. The probes consisted of a 4 cm diameter and 8 cm long aluminum chamber fixed to the end of a narrow 54 cm long stainless-steel conduit tube. The input and output optical fibers lead from the laser source and photodetector through the conduit tube to the Si thermometer chip. The silicon thermometer chip is mounted on an invar metal stage at the lower end of the chamber, using a spring clip to clamp the Si chip at a point several millimeters away from the ring thermometer. This mounting procedure allows the chip to freely expand and contract with temperature, and thereby avoid stress induced refractive index changes in the ring waveguide due to the stress-optic effect [13]. Mounting using rigid clamps or adhesives such as epoxy creates local stresses due to thermal expansion mismatch between the stage and clamps, adhesive layer, and the Si chip. Adhesive related stresses and creep have been observed to alter the thermometer reading by up to 170 mK over time and temperature cycling [2].

**Table 1. Ring Thermometer Properties**

| Ring | Cavity Length (mm) | Resonance Wavelength (nm) | Resonance Width (3 dB) (pm) | Resonance Depth (dB) | Directional Coupler Transmission* $|t|^2$ | Waveguide Loss in Ring* (dB/cm) |
|---|---|---|---|---|---|---|
| Ring A | 100 | 1554.605 | 12 | -16 | 0.992 | -2.42 |
| Ring B | 950 | 1553.514 | 9  | -11 | 0.967 | -2.75 |
| Ring C | 70  | 1535.807 | 15 | -23 | 0.994 | -3.36 |
| Ring D | 950 | 1557.980 | 8  | -13 | 0.968 | -2.32 |

* Experimental values determined by fitting the ring transmission function to the measured ring spectrum.

In the thermometer probes for Ring C and D, light was coupled to and from the chip using two polarization maintaining (PM) optical fibers mounted in a single 8° angle polished fiber block. The input and output fibers are separated by 500 μm spacing to match the on-chip grating coupler positions. The fiber block is aligned with the two grating couplers, and is permanently fixed in the probe with the fiber facets at a height of 250 μm above the chip as shown in Fig. 2.

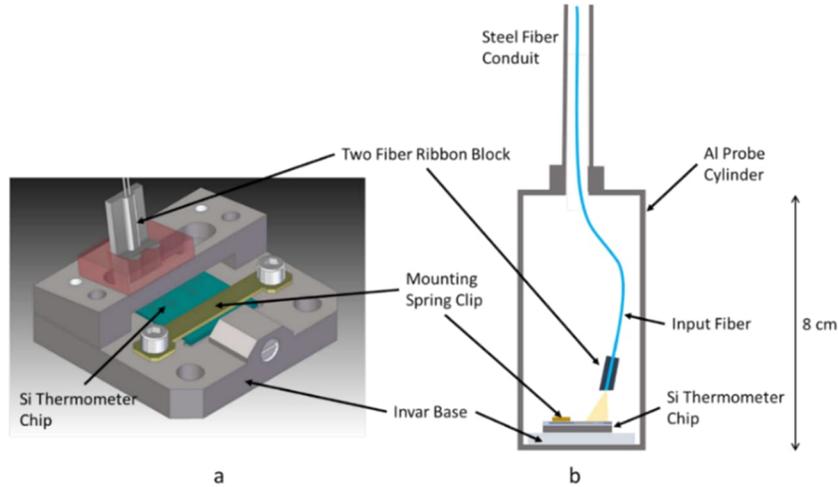

**Fig. 2.** The sealed thermometer probe assembly showing (a) a detailed view of the invar mounting stage and two fiber block assembly, and (b) a schematic side view cross-section of the entire probe (not to scale). The output fiber is parallel to the input fiber as shown in (a), but only the input fiber is shown in this cross-sectional view. The polished fiber block face is positioned 250 μm above the Si chip surface.

Positioning the fiber end facet well above the chip produces a diverging beam with an approximately 100 μm spot diameter at the chip surface. Although this results in a -15 dB mode mismatch loss from the fibers to the 10 μm wide grating couplers, this free space coupling system eliminates sensitivity to optical misalignment during probe assembly and also to dynamic misalignment due to thermal expansion when in use. The need for epoxy to bond fiber to the chip is also eliminated. The signal was still more than sufficient to make accurate resonance wavelength determinations. Furthermore, even in this configuration incident laser powers in the 1 mW range generated measurable resonance wavelength changes due to self-heating.

*2.2 Optical and Temperature Measurement*

The variation of resonance wavelength with temperature is monitored by repeatedly scanning the wavelength of light transmitted through the ring resonator across the selected resonance line as the chip temperature is varied. Accurate ring thermometer assessment therefore depends on the stability and accuracy of laser wavelength and reference thermometer calibrations, as well as precise temperature control with a negligible temperature gradient between the reference thermometer and ring.

For the bench-top stage measurements on Rings A and B, the silicon chip was placed on a temperature-controlled copper block mounted on a thermoelectric cooler (TEC) with the upper chip surface exposed to the laboratory ambient environment. This bench-top stage provides more flexibility in carrying out optical testing compared with the temperature calibration bath, for which the ring and optical fibers are be permanently mounted in a sealed probe assembly. However, the stage temperature accuracy and reproducibility are limited to approximately ±0.05 K by the stability of the TEC controller, fluctuating room environment, and the thermal gradients between the underlying stage and the exposed silicon-on-insulator chip surface [7]. As a result, the correspondence of the measured resonance wavelength to a specific temperature has a much larger uncertainty of ±0.05 K (or ±4 pm). The stage temperature was measured using a thermistor embedded in the stage, and the thermistor signal was used as feedback to the TEC controller to maintain the stage at a desired fixed temperature point. A one-dimensional heat flow model [7], that includes thermal radiation effects, predicts that the difference between stage temperature and Si waveguide temperature will be less than 50 mK for an ambient room temperature near 20 °C and stage temperatures between 20 °C to 50 °C. Light from an Agilent 81600B tunable laser is coupled into a polarization controller followed by a polarization maintaining (PM) single mode fiber with an 8° angle polished facet, which directs light to the input grating coupler on the chip. A second PM fiber captures light from the output grating and directs it to the photodetector. For the stage measurements on rings A and B, the two fibers were positioned directly above and in near contact to the grating couplers at an 8° incident angle.

The temperature control is much more precise for the Ring C and D temperature calibration bath measurements. The entire probe was immersed at a depth of 36 cm in a stirred liquid temperature calibration bath. The temperature of the bath was measured using a calibrated standard platinum resistance thermometer (SPRT) placed beside the probe, and two calibrated thermistors. The SPRT thermometer was used as the reference thermometer for ring measurements. By varying the position of the thermistors, the bath temperature uniformity of better than 0.4 mK was measured, while the temperature stability over time at any given set point was better than 0.7 mK [14]. The ring transmission spectra for the probe mounted Ring C and Ring D devices were measured using an HP model 81682A tunable laser and a Keysight 81624B photodetector. The laser is able to scan wavelength from 1480 nm to 1580 nm with a minimum wavelength step size of 0.1 pm. The wavelength of each data point was obtained directly from the laser readout. The laser calibration and repeatability were assessed by using the laser to measure the absorption lines of HCN and acetylene gas reference cells. The wavelength reproducibility from run to run was within ±0.1 pm, with a scatter in wavelength offsets within ±0.25 pm of the published reference gas line positions [15,16]. Given the measured resonance wavelength shifts near 77.5 pm/°C in these rings, the temperature measurement error arising from the laser wavelength uncertainty will be less than 3 mK. The details of the calibration procedures are reported in a separate publication [14].

## 3. Ring resonator model

A phenomenological ring model was developed in order to aid in understanding the potential thermometer reading error in a ring caused by optical self-heating and other optical nonlinearities, and also by the presence of back-reflections and fluctuations in incident light power. The model is based on the fundamental ring resonator equations as given by Yariv [17] and Bogaerts et al. [18]. In order to reproduce the effect of self-heating and other intensity dependent changes, the model includes a power dependent waveguide effective index. In the all-pass configuration of Fig. 1, the optical power $P_T$ transmitted through the through-waveguide is given by

$$P_T = P_0 \cdot \frac{a^2 + |t|^2 - 2a|t|\cos\psi}{1 + a^2|t|^2 - 2a|t|\cos\psi} \quad (1)$$

where $P_0$ is the incident power coupled to the through waveguide, $t$ is the field transmission coefficient of the directional coupler, and $a$ is the ring round trip field attenuation. The optical power circulating inside the ring is

$$P_{circ} = P_0 \cdot \frac{a^2(1-|t|^2)}{1 + a^2|t|^2 - 2a|t|\cos\psi} \quad . \quad (2)$$

The total round trip phase delay $\psi$ of the ring resonator is a function of the wavelength $\lambda$ and waveguide effective index $N_{eff}$. This phase is the sum of the propagation phase delay through the ring waveguide of physical length L and any additional phase $\phi_t$ arising from propagation through the directional coupler section.

$$\psi = \frac{2\pi N_{eff}}{\lambda} \cdot L + \phi_t \quad . \quad (3)$$

It is assumed that $\phi_t$ is only weakly dependent on wavelength and can be treated as a constant over small wavelength intervals comparable to the ring resonance linewidths. The transmission resonances occur at wavelengths where $\psi$ is a multiple of $2\pi$, and manifest as periodic minima in the transmission spectrum $P_T$. The depth of the resonance minima will depend on the values of ring round trip loss $a$ and the coupling coefficient t. At the critical coupling condition $|t| = a$, the transmitted power $P_T$ is zero on resonance. The ring waveguide effective index $N_{eff}$ is temperature dependent due to the thermo-optic effect in silicon, and hence resonance wavelength will shift with temperature, thereby enabling the use of the device as a thermometer.

The nonlinear component of the effective index in the model, $\Delta N_{eff}$, is a function of the circulating power in the ring so that $N_{eff} = N_{eff,0} + \Delta N_{eff}(P_{circ})$, and it follows that the transmission, phase, and circulating power in Eq. 1-3 will all vary with the incident optical power. Given a specific functional form for the $\Delta N_{eff}(P_{circ})$, the circulating power and transmission for the ring can be calculated by finding the values of $P_{circ}$ that self-consistently solve Eq. 2 for any specific incident power $P_0$. The resulting $P_{circ}$ and $N_{eff}$ can then be used in Eq. 1 to obtain the ring transmission spectrum $P_T$. This procedure will generate up to three different solutions depending on the wavelength, nonlinearity, and incident power. Fig. 3 illustrates the qualitative resonance line shapes expected for Si ring resonator at three different incident power levels, assuming a linear power dependent effective index, as would occur for simple self-heating caused by linear absorption. The transmission spectrum for zero incident power $P_0 = 0$ (equivalent to setting $P_{circ} = 0$ or $\Delta N_{eff} = 0$) is given by Eq. 1 and Eq. 3 directly. For low incident power, the light induced change in effective index causes the observed resonance line shape to become asymmetric and the resonance minimum to move to longer wavelengths. At higher power in Fig. 3, there are three possible solutions for $P_{circ}$ and the corresponding ring transmission $P_T$. In this region, the Si ring can exhibit bistability and self-oscillation. The nonlinear optics of Si resonators, and the nonlinear mechanisms such as self-heating, free carrier generation, and the intrinsic intensity dependent refractive index n₂, has been the topic of extensive previous and ongoing research [11, 19-22]. Temperature determination would use the lowest possible power that gives an adequate transmission signal with minimum possible nonlinear effective index change and local heating. Therefore, this work focuses only on the low power regime where the line shape is stable and repeatable, but temperature reading errors may still be caused by an optically induced resonance wavelength shift and also the line shape asymmetry, which can skew the results of line shape fitting procedures.

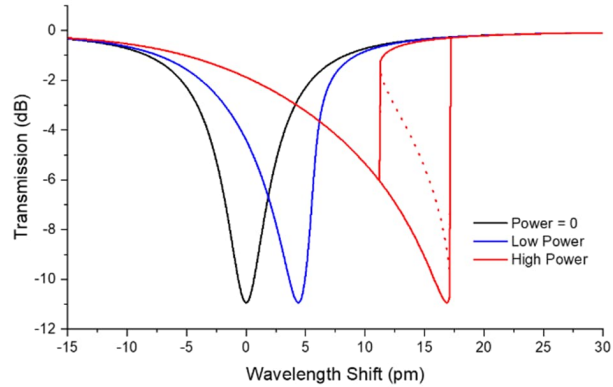

**Fig. 3.** The qualitative shapes of a Si ring resonance expected for zero power, low power and high input power, assuming a power dependent waveguide effective index, as would be the case for linear optical self-heating. The wavelength scale is referenced to the position of the resonance at zero incident power. At high power the ring transmission exhibits three possible solutions between $\lambda$= 12 pm and 17 pm, where the ring transmission will exhibit bistable behaviour.

The resonance line shape asymmetries shown in Fig. 3, and observed in experiment, are in fact only apparent distortions that result when the data is acquired by scanning the incident wavelength. The intrinsic resonance does not change shape, provided that there are negligible optically induced changes in the ring round trip loss *a*, or the ring DC coupling strength. The resonance center wavelength simply adjusts relative to the instantaneous laser probe wavelength, changing at each wavelength step in response to the circulating power $P_{circ}$ to produce the observed transmission. But in the low power regime, the resonance center wavelength of the intrinsic ring resonance coincides exactly with the measured ring transmission minimum wavelength even though measured resonance shape appears to be distorted. Therefore, the resonance wavelength shift determined from the measured resonance minimum accurately reflects the actual ring effective index as determined both by the underlying chip temperature, and local self-heating and other nonlinear optical effects within the waveguide. In this power regime the measured resonance minima wavelengths can be compared with predictions of theoretical models for the induced refractive index changes. On the other hand, if incident power is high enough to push the ring into the bistable regime, this is no longer the case since the observed transmission spectrum may be a time average of the fluctuations between possible resonator states. The use of ring resonators as thermometers or other types of sensors therefore requires consideration of the incident power and possible nonlinearities, particularly for very large quality factor (Q) resonators where the stored circulating power may be orders of magnitude larger than the incident beam power. In the ideal thermometer, accuracy and repeatability are independent of the instrumentation and the laser power used to interrogate the device. Since this is not possible in a real device, the goal of this work is to quantify the magnitude of the potential errors and explore means to improve accuracy.

## 4. Self-heating and Power Dependent Effective Index Changes

The transmission spectra for Rings A and B were acquired at different optical powers using the bench top optical stage maintained at a temperature of 25 °C ±0.05 using the TEC controller. Fig. 4a shows the measured spectrum of Ring A at incident power levels from 50 µW to 2 mW. Far from resonance and at 1 mW (0 dBm) input laser power, the received power at the photodetector is -15 dBm, including approximately -1 dB loss from the input polarization controller. Since the ring is positioned at the approximate mid-point of the optical path from laser to detector, the accumulated loss in the through waveguide up to the position of the ring is estimated to be –7 dB, or 20% of the launched laser power. The losses include a grating coupling efficiency of –5 dB from fiber to the Si waveguide, Si waveguide loss of approximately 1.5 dB/cm ±0.5 and any additional system losses between laser and detector. For simplicity the input powers given in Fig. 4 are as launched into the input fiber from the laser.

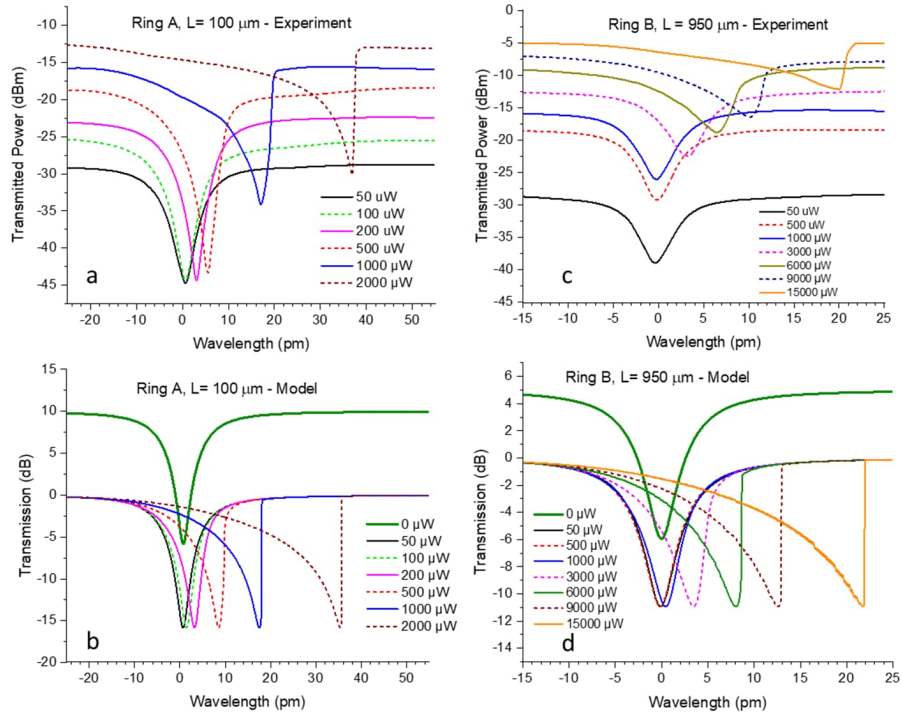

**Fig. 4.** (a) The measured and (b) calculated transmission resonance spectra for Ring A at input laser powers from 0 µW to 2000 µW, showing resonance wavelength shifts and line shape distortion resulting from optical power induced effective index changes. The measured and calculated resonance spectra for Ring B at laser powers up to 15 mW are shown in plots (c) and (d) respectively. The chip temperature is held at T = 25 °C ±0.05. The nominal resonance wavelengths are given in Table 1, but to facilitate comparison of the four graphs, the wavelength scale is shifted so that λ = 0 pm corresponds to the line centre at zero incident power. The calculated line shapes for 0 µW power are vertically offset for clarity. The empirical power dependent effective index derived from the Ring C measurements is used to produce the model results shown in b) and d).

The measured spectra of Fig. 4(a) and 4(c) show that the ring resonance wavelengths are shifted to longer wavelengths with increasing power and the asymmetric line shape distortion is already evident in Ring A at incident powers as low as 200 µW for Ring A. At highest powers the sharp transition at the long wavelength edge of the resonance indicates that the rings are entering the bistable regime and undergo a switching transition. This occurs near 1 mW for Ring A, and above 10 mW for the longer cavity in Ring B. The corresponding model results in Figs. 3(b) and 3(d) were generated using an empirical nonlinear index function determined by fitting to spectral measurements of the Ring C thermometer carried out in the temperature calibration bath, as described below. The model calculations for each ring used the DC through coupling coefficients and ring waveguide losses in Table 1, which are derived from the measured -3 dB linewidths and resonance depths for each ring. Fig. 5 shows the measured and calculated power induced resonance wavelength shifts corresponding to the spectra in Fig. 4. Both the model line shapes and wavelength shifts are in good agreement with experiment. Ring A shows a wavelength shift of 15 pm for a 1 mW input power. When used as a thermometer, this generates a reading error of almost 0.2 K based on resonance minimum wavelength. The distortion of line shape will cause additional ambiguity if line shape fitting is used to determine the resonance position. The optically induced wavelength shift for Ring A in Fig 4a is more than ten times larger than for Ring B at any given power. This result is expected since for rings with similar quality factors or resonance line width, the circulating power in a ring at resonance will scale inversely with cavity length [22], as will be discussed in more detail at the end of this section.

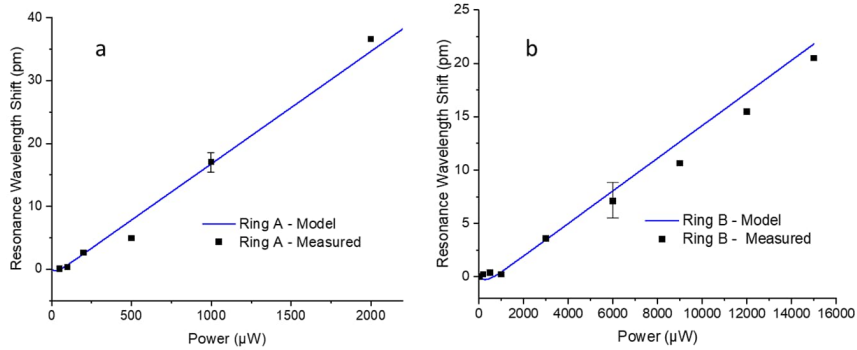

**Fig 5.** The measured and calculated resonance wavelength shift with incident power for (a) Ring A and (b) Ring B. Resonance wavelengths are determined from the minimum points of the resonance lines shown in Fig. 4. The chip temperature is held at T = 25 °C. The stage temperature stability of ±0.05 C generates a measured resonance wavelength uncertainty of approximately ±3 pm due to temperature induced resonance drifts.

Similar measurements of the resonance wavelength variation with optical power were carried out for Ring C and Ring D, and the results are shown in Fig. 6. Since laser to chip coupling is much lower in the bath experiments, the powers given here for the Ring C and D have been scaled to correspond to the incident power scale used for Ring A and Ring B in Figs. 4 and 5 to allow a direct comparison of all the ring devices. The probes were immersed in the temperature calibration bath maintained at a constant temperature at T = 23 °C. The temperature bath stability of ±0.7 mK is much better than that for the bench-top stage used for the Ring A and B measurements, so the resonance wavelength uncertainty in Fig. 6 is dominated by the ±0.1 pm repeatability of the laser, and by the 0.1 pm wavelength step size used to acquire the spectra. Each data point in Fig. 6 is an average over approximately ten separate measured spectra, and the error bars shown in Fig. 6(b) indicate the typical standard deviation of each data set. The Ring C resonance wavelengths in Fig. 6a show a positive linear change with power for coupled laser powers higher than 50 µW, similar to the behaviour of Rings A and B. At very low power Ring C displays an initial blue shift of the resonance, suggesting the presence of a second negative optically induced index change mechanism that competes with the positive index change that dominates at higher powers. The power induced wavelength shift for Ring D also shows a comparable small blue shift, but with the power dependence that is an order of magnitude slower than that for Ring C. In both devices the maximum blue shift is approximately -0.2 pm. In Ring D the measured wavelength shift never reaches the positive linear regime at the input powers available in the probe experiments. Again, this is expected because the cavity length of Ring D is almost ten times longer than in Ring C.

The power dependence of the resonance wavelength in all four devices is a result of light induced effective index changes in the silicon waveguide. The contributions to optically induced index changes in semiconductors will include self-heating due to absorption of light in the waveguide. As will be discussed below, this can occur even for photon energies lower that the band gap. There will also be direct refractive index contributions arising from optically excited free carriers, as well as band-filling and absorption saturation due to free carrier excitation and defect state filling, and the intrinsic intensity dependent refractive index $n_2$ [11, 23-25]. An empirical model for the intensity dependent waveguide effective index was developed to aid in comparing the experimental results for the four ring devices. The functional form of the nonlinear index $\Delta N_{eff}$ was chosen by using ring resonator model for the geometry of ring C, and selecting a function and parameter values that generate a good match to the experimental data of Fig. 6. The resulting equation has the form

$$\Delta N_{eff} = A \cdot P - R(1 - e^{-\gamma P}), \tag{4}$$

with coefficients $A = 1.39 \times 10^{-6}$ mW$^{-1}$, $R = 3.01 \times 10^{-6}$, and $\gamma = 1.14$ mW$^{-1}$ where the in-waveguide optical power P is expressed in milliwatts. The uncertainty in these empirical parameters is dependent on the uncertainty in power attenuation between the laser up to the position of the ring. The form and parameter values for Eq. 4 are based on the data for Ring C in Fig. 5, but the same index function also generated excellent matches to the measured spectral data for Rings A, B, and D with no additional adjustments, as shown by the calculated curves in Figs. 5 and 6. This suggests that Eq. 4 is generally applicable to Si waveguides comparable to Ring C at the low powers used in these experiments. Although not clearly evident in the experimental data, the small negative index changes in Ring C and Ring D at low power (i.e., the resonance blue shift) are also likely present in Rings A and B, but are not resolved because of the larger temperature fluctuations of the bench-top stage.

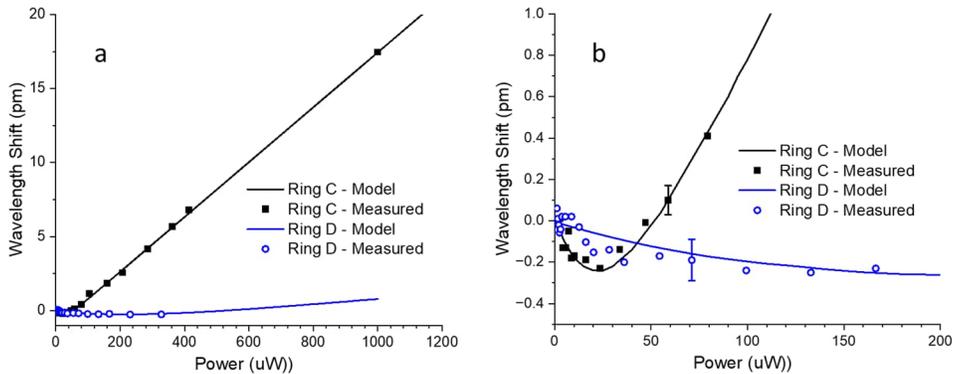

**Fig. 6.** (a) The measured and calculated resonance wavelength variation with incident power for Ring C (L= 100 μm) and Ring D (L= 950 μm). The rings are maintained at T = 23 °C with less than ±0.001 °C variation during the measurements. (b) An expanded view of the low power range. The wavelength axis origin (Δλ= 0 pm) corresponds to resonance wavelengths of λ=1535.807 nm and λ=1557.981 nm for rings C and D respectively.

The first term in Eq. 4 describes a positive linear power dependence that could be attributed to either self-heating by linear optical absorption processes or the intensity dependent refractive index $n_2$. The measurements suggest that self-heating dominates, since value of $n_2 = 5.6 \times 10^{-18}$ m$^2$W$^{-1}$ for silicon [23] is more than an order of magnitude too small to account for the empirically derived $\Delta N_{eff}$. The photon energy of the λ=1550 nm light (~ 0.8 eV) used in these experiments is much lower than the Si indirect and direct band gap energies (~ 1.1 and 3 eV, respectively). Linear absorption and linear electron and hole carrier excitation are therefore not expected in bulk Si at wavelengths longer that λ = 1200 nm. However, carriers may still be generated via mid-gap defect and surface states. The photonic wire waveguides used in this work have a large surface to volume ratio, and fabrication processes such as reactive ion etching can introduce a high concentration of near surface defects states. Ion implanted defects, as would be created by reactive ion etching, are known to produce linear absorption in silicon [26]. Several studies have found that linear absorption by surface states and near surface defects can dominate optical absorption at low powers [11, 26-30] in Si photonic waveguide devices, and therefore contribute to heating and provide a linear free carrier excitation path via mid-gap states. Novarese et al. have found that including sub-band gap energy linear absorption is essential to explain their experimental results on nonlinear ring resonator response [11].

The second term in Eq. 4 models the small negative index change responsible for the blue shift in Fig. 6, which appears to level off with increasing power. A negative index change in a semiconductor material may arise from optically excited free carriers, or possibly from defect state filling and/or band filling [11, 25]. Similar behaviour to that of ring C has been observed in previous work, and was attributed to the changing ratio of a negative and positive index changes induced by free carrier excitation and self-heating respectively [11]. The 0.2 pm blue shift observed here corresponds to an effective index change of approximately $\Delta N_{eff} = 10^{-6}$, which would result from a very small electron and hole density change of the order of $\Delta n, \Delta p \sim 10^{14}$ cm$^{-3}$. In the analysis presented in Ref. [11], the excited carrier lifetimes are long at low carrier densities, so that carrier density increases rapidly and the carrier induced index change dominates at very low optical powers. As power and hence carrier density increase, the free carrier lifetime becomes shorter causing carrier energy to be more rapidly dissipated as heat while suppressing the growth of carrier density. Higher order absorption processes such as two photon absorption (TPA) may also contribute to the effective index change via heating and carrier excitation. However, the experimental results in Figs. 4 and 5 show no obvious quadratic wavelength variation with power. This result is consistent with the thermal modeling calculations of Dickmann et al. that predict the threshold for TPA heating to be near 10 mW input waveguide power, for a ring of comparable dimensions to Ring C [22]. After adjusting for our estimated coupling efficiency, this is well above the input waveguide powers in our measurements. Note that the index model of Eq. 4 is a purely phenomenological model intended to predict and compare ring thermometer behaviour under different measurement conditions on the assumption that similar results will be obtained for comparable devices produced by commonly used Si fabrication processes. Identifying the underlying physical mechanisms responsible for the observed silicon waveguide index change can only be fully addressed using time resolved optical experiments in combination with detailed modelling of carrier dynamics and heat flow, and is outside the scope and objectives of this study.

The power induced resonance changes for Rings A and C are approximately 15 pm for a typical interrogating laser power of 1 mW. Under these operating conditions, a temperature reading error of more than 100 mK would result when using a ring as a thermometer. This would preclude using Si rings for demanding measurement applications and as metrology reference standards. An obvious solution is to correct every measurement for the intensity dependent error using a calibrated correction factor for laser power. This approach requires a calibration of the ring thermometer response to different laser powers, and precise control of laser power and laser coupling to the Si chip. The power dependent resonance shift can also be reduced simply by using the lowest possible optical power that gives a sufficient signal to noise ratio to accurately determine resonance positions. In our experiments, incident powers near 10 μW were sufficient to determine resonance positions to within ±0.1 pm. At this level the power induced wavelength shift of ring C is approximately 0.2 pm, corresponding to a temperature error of less than 3 mK.

It would be more attractive to introduce ring design modifications that reduce the intrinsic power dependence of the resonance positions, so that accuracy and repeatability can be decoupled from the details of the measurement instrumentation. The results for Ring B and D in Fig. 5 and Fig. 6 show that simply increasing ring length can give a significant improvement.

Ring B and D have a length L= 950 μm which is approximately ten times longer that both Rings A (L= 100 μm) and C (L= 70 μm). The corresponding power induced resonance shift is more than ten times smaller than that for Rings A and C, while still having a comparable narrow resonance linewidth $\Delta\lambda_{3dB}$. This is in agreement with basic ring resonator theory that predicts that the circulating power will decrease inversely with cavity length when $\Delta\lambda_{3dB}$ or Q-factor stays the same [22]. When the ring loss and the directional coupler coupling coefficient are small, the circulating power given by Eq. (2) can be approximated as

$$P_{circ} \sim P_0 \cdot \frac{2\lambda^2}{\pi N_g} \cdot \frac{1}{\Delta\lambda_{3dB} L} = P_0 \cdot \frac{2 \cdot FSR}{\pi \Delta\lambda_{3dB}} \quad . \tag{5}$$

where FSR is the free spectral range of the ring. A narrow linewidth $\Delta\lambda_{3dB}$ is desirable to improve the precision of resonance wavelength determination, but simply optimizing the ring design to narrow the linewidth increases the circulating power. By adjusting the ring DC coupling coefficient as the ring length L is increased, the ring line width can be kept narrow while reducing the circulating power and hence self-heating and other optical nonlinear effects. The maximum usable ring length will eventually be limited by the increased ring round trip loss, which will cause linewidth broadening that can no longer be compensated by reducing the directional coupler coupling coefficient.

The variation of the power induced resonance wavelength shift with ring length is illustrated in Fig. 7, which shows the calculated nonlinear response for several model rings of different cavity lengths but all with the same power dependent effective index of Eq. 4. The waveguide loss was assumed to be 2 dB/cm in each case, and the directional coupler coefficients were adjusted so that each ring was critically coupled. The resonance 3 dB linewidths of $\Delta\lambda_{3dB} = 8.4$ pm were approximately the same for each ring.

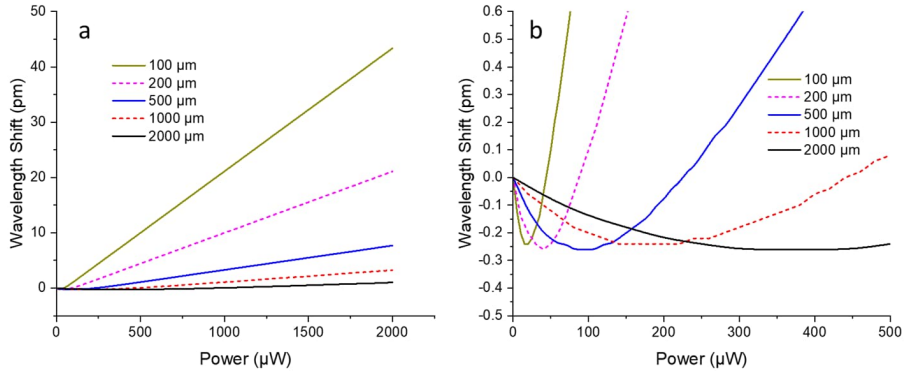

**Fig. 7.** (a) Comparison of power induced wavelength shifts for Si ring resonators of various lengths from 100 μm to 2000 μm, using the effective index model of Eq. 4. (b) An expanded view showing behaviour at low power. In these calculations the waveguide loss was fixed at 2 dB/cm, and the ring directional coupler was adjusted to produce critical coupling at each ring length.

The trends illustrated in Fig. 7 are confirmed by the experimental results presented in Fig. 5 and 6. Comparing rings with L= 100 μm and L= 2000 μm in Fig. 7a at input powers of 1 mW, the strong red shift from self-heating can be reduced from 20 pm to less than 0.2 pm. The small initial blue shift is more difficult to avoid since it occurs at very low powers. However, for input powers less than 100 μW, the power induced change is suppressed to less than 0.1 pm, corresponding to temperature reading errors of less than 1 mK. For most applications this is negligible, but may be important for demanding metrology applications. The disadvantage of extending cavity length is that the FSR decreases with cavity length, limiting the temperature change before adjacent order resonances overlap. For example, the FSR in rings B and D is only 0.6 nm so that an 8 °C temperature change will cause the resonances to shift by the full FSR. This leads to a temperature reading ambiguity if the ring is interrogated intermittently, since the resonance order will not be known if a temperature change of more than 8 °C occurred in the interval between measurements. On the other hand, the ambiguity can be avoided if resonances are tracked continuously in real time.

## 5. Optical scattering and back reflection

### 5.1 Intra-ring scattering: Resonance line splitting and distortion

Stray light from unwanted reflections and scattering is present to a greater or lesser degree in all integrated optical chips, and can cause significant changes in performance for integrated photonic devices. In this section, we examine the effects of back-reflections and scattering specifically in a ring resonator thermometer, but the considerations are equally relevant to many types of resonator-based sensors. Light that is scattered out of the waveguide into the substrate or cladding simply contributes to the total waveguide loss. While higher loss will increase the resonator linewidth, loss alone will not otherwise change the ring behaviour. On the other hand, light that is scattered either once or many times into counter-propagating waveguide modes (i.e., back reflections) can cause significant changes in the device output. Within a photonic chip every transition between different waveguide structures will produce some back reflection. Light back scattered from sidewall roughness and defects into the waveguide also contributes to the stray light propagating through the system, and can be regarded as a distributed back-reflection.

Back scattering within the ring resonator has the most serious and obvious effect on ring output. Resonance line splitting up to several picometers and asymmetry in overall line shape are commonly observed in Si waveguide micro-resonators [27, 31-33] and are also present in the ring resonators used in this work. These line shape distortions create ambiguities in identifying a precise resonance wavelength, and may introduce temperature further reading errors if the distortions are wavelength or temperature dependent. Back-scattering within micro-resonators has already been described by many groups [27, 31-35]. Here the basic phenomenology and descriptive equations are only briefly summarized, in order to better understand the impact of line shape distortions on temperature reading accuracy and repeatability.

Strong back-scattering by distributed sidewall roughness or isolated defects within the ring cavity will couple light from one ring mode into the degenerate counter-propagating ring mode of the same order. The two coupled modes combine to form two resonances separated in wavelength and the observed resonance line shape in the ring output spectrum splits into two closely spaced resonance minima. Back scattering and reflections in the directional coupler (DC) can also excite the counter-propagating waveguide mode by directly coupling the incident beam backwards into the ring [32]. Such back coupling arises from reflections at the two ends of the DC if the through waveguide and ring junctions are not sufficiently adiabatic, and also from any sidewall roughness along DC parallel waveguide section. Back coupling can be comparable to in-ring back-scattering and cause strong line shape asymmetries even when the underlying scattering mechanisms are wavelength independent [32]. The resulting distortions become increasingly difficult to avoid as resonator Q increases, since the counter-propagating mode power becomes larger when the total ring loss is low compared with the back scattering strength. An additional complication is that the line shape distortion can be strongly wavelength dependent. Even for the same ring the individual ring modes at different resonance orders may exhibit different Q-factors, mode splitting, and wavelength offsets from the nominal resonance position [27,31,32]. One resonance may have a perfect single peak Lorentzian line shape as expected from Eq. 1, while another resonance only a few nanometers away may be split into two peaks, as is evident from the measured line shapes in Fig. 8.

Resonator mode splitting is usually treated using time dependent coupled mode theory [32-37]. Coupled mode theory has been applied in many previous publications, but we repeat a brief outline here in order to arrive at a descriptive form for the all-pass transmission, which has not been the focus of previous work but is the transduction signal in our thermometers. The excitation and coupling of two counter-propagating ring mode field amplitudes $A_+(t)$ and $A_-(t)$ can be described in terms of an internal ring back scattering strength β, and the directional coupler forward and back coupling coefficients, $κ_f$ and $κ_b$, respectively. The coefficient β lumps together both distributed sidewall roughness scattering and any scattering from specific defects in the ring. It is assumed that the input field $S_{in}·exp(iωt)$ has an optical frequency ω and the circulating modes therefore have a time dependence of form $A_±(t)·exp(iωt)$. The driving frequency ω is also assumed to be close to the nominal frequency $ω_0 = m\,(2πc/N_{eff})·L$ for the resonance of order m.

$$\frac{d}{dt}A_+(t) = -(i\Delta\omega + \delta_0 + \delta_c) + i\kappa_f S_{in} + i\beta A_-(t) \qquad 6(a)$$

$$\frac{d}{dt}A_-(t) = -(i\Delta\omega + \delta_0 + \delta_c) + i\kappa_b S_{in} + i\beta A_+(t) \qquad 6(b)$$

Here $S_{in}$ is the input waveguide mode field amplitude, $\Delta\omega = \omega - \omega_0$ is the frequency offset from resonance, $\delta_0$ is the exponential attenuation coefficient due to intrinsic waveguide loss and $\delta_c$ is the corresponding mode attenuation due to coupling of light from the ring into the output waveguide by the directional coupler by backwards and forward coupling. Setting both Eq. 6(a) and 6(b) to zero yields the time independent expressions for the two counterpropagating mode fields within the ring.

$$A_- = \frac{\kappa_b(i(\delta_0+\delta_c)-\Delta\omega)-\kappa_f\beta}{(i(\delta_0+\delta_c)+\Delta\omega)^2+\beta^2}\cdot S_{in} \qquad 7(a)$$

$$A_+ = \frac{\kappa_f(i\delta_0+i\delta_c-\Delta\omega)-\kappa_b\beta}{(i\delta_0+i\delta_c+\Delta\omega)^2+\beta^2}\cdot S_{in} \qquad 7(b)$$

The circulating power in these modes will have two maxima at $\omega_\pm = \omega_0 \pm (\beta^2 - (\delta_0+\delta_c)^2)^{1/2}$ if the back scattering coefficient β is larger than the total loss $\delta_c + \delta_0$ [33]. For smaller values of β there is no splitting and transmission minimum occurs at the nominal resonance frequency $\omega = \omega_0$, i.e., at $\Delta\omega = 0$. In the all-pass configuration used in the thermometer devices, the final output field in the through waveguide, $S_{out} = S_{in} + i\,\kappa_f^* A_+ + i\kappa_b^* A_-$, is a superposition of contributions by the incident field, the desired forward coupled $A_+$ mode and unintentional backwards coupled counter-propagating $A_-$ mode [34]. Using the square modulus of the output field $S_{out}$, the ring power transmission can be arranged into the form

$$T_{out} = \frac{|S_{out}|^2}{|S_{in}|^2} = \frac{1}{(\Delta\omega^2+(\delta_0+\delta_c)^2-\beta^2)^2+4\beta^2(\delta_0+\delta_c)^2}$$
$$\cdot\left\{\left[\left(\Delta\omega^2+\delta_c^{\,2}-\delta_0^{\,2}-\beta^2\right)^2+4\delta_0^{\,2}\Delta\omega^2\right]+\Gamma^2\beta^2\left[1-\frac{4\delta_0}{\Gamma\beta}\cdot\Delta\omega\right]\right\} \qquad (8)$$

The denominator in Eq. 8 has a similar form to that of $|A_+|^2$ and $|A_-|^2$ with minima at the split resonance frequencies $\omega_\pm$ [33]. The first part of the numerator is symmetric in $\Delta\omega$, while the second part is asymmetric and depends on the coefficient Γ,

$$\Gamma = \kappa_b^* \kappa_f + \kappa_b \kappa_f^* = 2|\kappa|^2\,|f|\cos\phi_f, \qquad (9)$$

which varies with the complex ratio of back and forward coupled field strengths, $f = k_b/k_f$, as introduced in Ref. 32. Here $\phi_f$ is the phase of f. Equation 8 predicts different line shape behaviors depending on the values of β and $κ_b$. In the simplest case the back scattering and DC back coupling are assumed to be wavelength and temperature independent. In the absence of any back scattering in the ring (β = 0) or back coupling in the directional coupler ($κ_b$=0), Eq. 8 gives the same spectral line shape predicted by Eq. 1. If back scattering in the ring is present but below the threshold for mode spitting (β < $δ_c$ + $δ_0$), the transmitted line shape remains symmetric and retains a single minimum at $ω_0$, and the transmitted power is zero at the critical coupling condition $δ_c^2 = δ_0^2 + β^2$ [26, 32-34]. Back scattering will cause some line broadening, but in the weak scattering regime the resonance minimum remains at the nominal resonance frequency $ω_0$ (or wavelength $λ_0$), and will shift with temperature at the same rate as in an ideal silicon ring. Therefore, weak internal ring scattering does not impair the accuracy or practical viability of a ring photonic thermometer. Fig 8(a) shows an example of a line shape typical of the low back scattering regime, measured for Ring D. When the internal ring back scattering exceeds the threshold for mode splitting (β > $δ_c$ + $δ_0$) the internal power circulating in the ring has maxima at the two frequencies $ω_±$, and the transmitted spectrum splits into two minima of equal depth. The more complex functional form of the numerator in Eq. 8 means that the frequencies of the transmission minima differ slightly from the peaks of the internal mode of Eq. 7, but will still be of equal depth and equidistant about the nominal resonance center frequency $ω_0$. Since $ω_0$ still changes with temperature in the same way as an ideal ring, the split pair average center frequency (or wavelength) would in principle remain an accurate indicator of ring temperature.

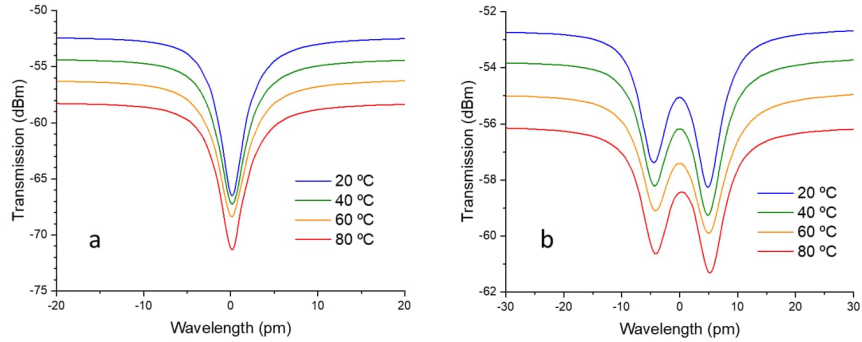

**Fig. 8.** (a) The measured transmission spectrum of a single lobed resonance for Ring D at four temperatures. (b) the split resonance spectrum at a different resonance order in the same ring. The resonances have been shifted to align the center wavelengths to l = 0 pm at each temperature and also offset vertically in order to facilitate a comparison of line shapes. The resonances are at l = 1558.3 nm and l = 1554.7 nm at t= 20 °c, for (a) and (b) respectively.

Finally, when the back-coupling in the directional coupler is included, then the last term in the numerator of Eq. 8 creates an asymmetry in the spectral line shape about the center resonance frequency. If mode splitting is present this will result in two asymmetric lobes in the transmission spectrum. This is the case for the measured resonance shapes in Fig. 8b. Even without obvious resonance splitting, an asymmetric distortion of the transmitted line shape may be present. It becomes impossible to determine the nominal ring resonance frequency $ω_0$ (or wavelength) and hence temperature without resorting to complex line shape fitting procedures. The concomitant complexity and ambiguity are obviously undesirable in a practical thermometer. Line splitting is often observed in our Si rings and can be up to 10 pm in wavelength separation as in Fig. 8(b), so the resulting uncertainty in resonance center wavelength could potentially limit temperature reading accuracy to more than 100 mK. The discussion so far is based on the simplifying assumption that back scattering and back coupling are wavelength independent. In reality the back scattering and back coupling can exhibit a significant wavelength dependence and the line shape distortions and resonance depths may be completely different from one mode order to the next for the same ring [31,32]. This wavelength dependence has been attributed to the statistical independence of the Fourier components of side wall roughness [31] and the interference effects when light is scattered repeatedly between different points in the ring and directional coupler [32]. This point is illustrated by the two measured resonances in Ring D shown in Fig. 8. These resonances are only 4 nm apart in wavelength yet the resonance in Fig. 8a has an almost perfect Lorentzian line shape while the line in Fig. 8b is broadened and split into a clearly separated and asymmetric pair. The line of Fig. 8(a) appears to be ideal for photonic thermometry or any other sensor based on resonance tracking. However, since scattering and back coupling are wavelength dependent, one might expect that as temperature and hence the resonance wavelength change, the line shape itself will evolve with temperature in an unpredictable way. This would render the device unusable as a thermometer because of the ambiguity in line fitting or peak finding if the line shape constantly evolves with temperature.

Fortunately, this is in fact not the case. We have found that ring resonance line shapes are surprisingly invariant over a wide temperature range. The single lobed Lorentzian mode of Ring D in Fig. 8a is shown at several temperatures from 20 °C to 80 °C. The line shapes at each temperature are indistinguishable from each other, even though the resonance center moves by more than 4 nm between T= 20 °C and 80 °C. For this resonance the scattering and back coupling effects appear to be negligible. Distortions do not appear as the temperature changes even though the temperature induced resonance wavelength change exceeds the 3.6 nm nominal wavelength difference between the two resonance orders at 20 °C shown in Fig 8a and Fig 8b. It is even more surprising that the strongly distorted resonance of Fig 8b is also completely invariant across the same 60 °C temperature range. The distorted line profile indicates that scattering and back coupling are obviously significant, but their relative magnitude and effect are completely unchanged by heating the ring. The explanation is that the scattered and back-coupled field propagation phases in the ring scale with the effective index of the waveguide $N_{eff}$, which changes with temperature due to the thermo-optic response of silicon. The main ring resonance mode phase also scales identically with $N_{eff}$. As temperature changes and the resonance wavelength (as measured in vacuum) changes, the physical wavelength $\lambda/N_{eff}$ in the ring cavity at the resonance center remains constant, as do relative phases of the fundamental ring mode fields, scattered fields, and back coupled fields. Hence their relative interactions are unchanged over a wide temperature range. On the other hand, the amplitudes of the electric fields scattered from specific point defects are temperature independent since the scattering depends on defect size and the index difference between the Si core (n ~ 3.5) and $SiO_2$ cladding (n ~ 2.0). This index step is much larger than any temperature induced index changes given the thermo-optic coefficients of $2\times10^{-4}$ °C$^{-1}$ for silicon and $1\times10^{-5}$ °C$^{-1}$ for $SiO_2$. Conceivably the line shape invariance may eventually break down over a very large temperature span because the device itself also undergoes thermal expansion. However, such deviations have not been observed over the temperature ranges used in this work. The conclusion is that if a resonance order with acceptable line shape can be identified, that specific resonance will remain unchanged and can be used for temperature measurement over a wide temperature range. In the Si rings used is this work, distorted and split resonances do appear frequently but the majority of resonance orders across the 1530 nm to 1580 nm working wavelength range have single lobed line shapes.

*5.2 Extra-ring scattering and back reflection: Background intensity noise*

Scattering and multiple reflections in waveguides external to the ring can also distort the ring line shape, and under certain conditions (e.g., strong coupling between the ring and reflection induced parasitic external resonant cavities) render the ring unusable as a sensor [39,40]. In this section, we assume that the back-reflections and back-scattering sufficiently small that only small perturbative changes in the resonance line shape occur. In a silicon photonic chip, the strongest reflections usually arise at the input and output waveguide facet couplers or surface grating couplers. Since the stray light can be back-reflected or scattered many times and interfere coherently with the main incident beam, reflections and scattering can produce both periodic ripple and random fluctuations in the intensity and phase of light propagating through even a simple straight waveguide. Significant progress has been made to design coupling structures that simultaneously maximize fiber to waveguide coupling efficiency and minimize back reflections [12,38], but given the sensitivity to small dimensional variations in as-fabricated features, reflections can never be completely suppressed even using optimal designs. Nevertheless, the effects on sensor performance can be minimized through the appropriate mitigation strategies.

As an example, Fig. 9a shows a measured spectrum for ring C at two different temperatures, for which a background intensity ripple is generated by input/output grating coupler back reflections of about R = 0.06. These couplers create a weak on-chip Fabry Perot (FP) cavity encompassing the ring, as shown schematically in Fig. 9(b). The cavity length between the two gratings is 2 mm. The induced intensity and phase fluctuations near the resonance can skew the apparent line shape and produce resonance wavelength measurement errors. Similar reflection effects can arise from elsewhere in the optical measurement train such as at optical fiber facets and at interfaces of any in-line optical components such as waveplates and polarizers. The effects of on-chip and off-chip reflections are similar, but on-chip sources of reflection cannot be eliminated once the chip has been fabricated.

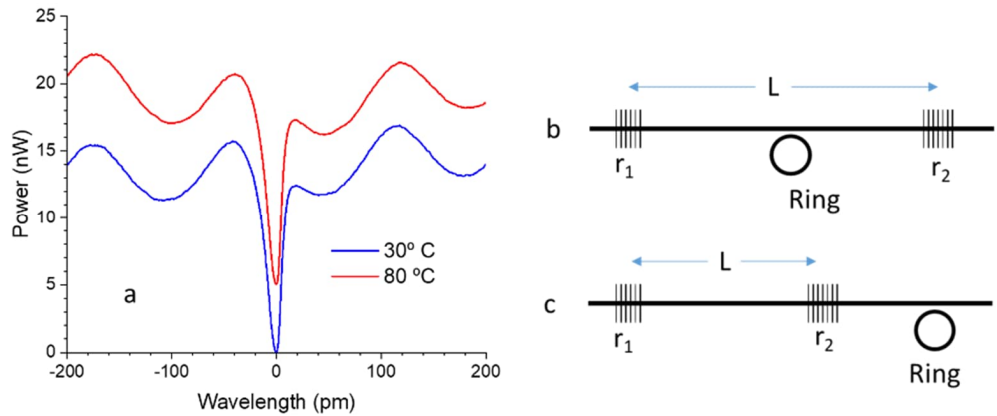

**Fig. 9.** (a) The measured transmission spectrum of Ring C at 30 °C and 80 °C showing a single resonance and the background intensity ripple due to back-reflections of approximately R= 6% from each of the input and output surface grating couplers. (b) Schematic diagram of back-reflection model with a ring resonator within the Fabry-Pérot cavity formed by two reflecting structures (e.g., two surface grating couplers as in Fig. 1), and (c) with the ring external to the Fabry-Pérot cavity.

We use the two simple models shown in Fig. 9(b) and Fig. 9(c) to examine the effect of reflections and back-scattering on thermometer reading accuracy. The first model consists of a ring coupled to the through waveguide and positioned inside the FP cavity formed between two waveguide-to-fiber couplers with a back-reflection coefficients $r_1$ and $r_2$. The other details of the couplers are not important in this discussion. This geometry forms a coupled FP cavity and ring resonator system in which the intensity and phase of the light in the through waveguide is modified by the ring on each FP cavity round trip. This model produces a background fringe pattern that is consistent with the measured Ring C spectrum shown in Fig. 9(a). In a typical photonic chip, the back-reflections are weak so that the total FP cavity loss is much larger than the ring DC coupling strength, and the perturbations in the transmitted ring resonance line shape will be small. The second model of Fig. 9(c) is similar to Fig. 9(b) except that the ring is now positioned outside the FP cavity. Here the intensity fluctuations are created by multiple reflections between two interfaces located on only one side of the ring, and therefore generate a simple external intensity background intensity modulation that is independent of the ring properties. Real devices may show more complex behaviour due to the superposition of several sources of back-reflection and scattering through the optical path, but these two models capture the basic physics and phenomenology of the two mechanisms that can distort the apparent resonance line shape, and can guide mitigation strategies to improve ring sensor measurement.

The transmission spectrum of the coupled ring-FP system in Fig. 9b can be calculated by including the amplitude and phase modulation of light due to the presence of the ring resonator in the internal propagation term of the FP transmission function [41]. The complex through transmission of the ring resonator directional coupler is

$$B_r = \frac{-a+te^{-i\psi}}{-at^* + e^{-i\psi}} \quad . \tag{10}$$

where the coefficients are defined as for Eqs.1-3 above. Inserting $B_r$ into the FP transmission equation gives

$$T_{FP} = T_0 \left| \frac{1}{1-B_r^2 r_1 r_2 e^{i\delta}} \right|^2 \tag{11}$$

The coefficients $r_1$ and $r_2$ are the field reflectivity of the FP mirrors (e.g., the input and output grating couplers), and $\delta = 4\pi N_{eff}/\lambda$ is the round-trip phase accumulated by through waveguide propagation within the FP cavity, but not including the phase modulation introduced by the ring by the ring transmission $B_r$. The factor $T_0$ incorporates the coupling efficiency and losses of the grating couplers. Here $T_0$, $r_1$ and $r_2$ are treated as a wavelength independent constant in the immediate neighbourhood of the ring resonance.

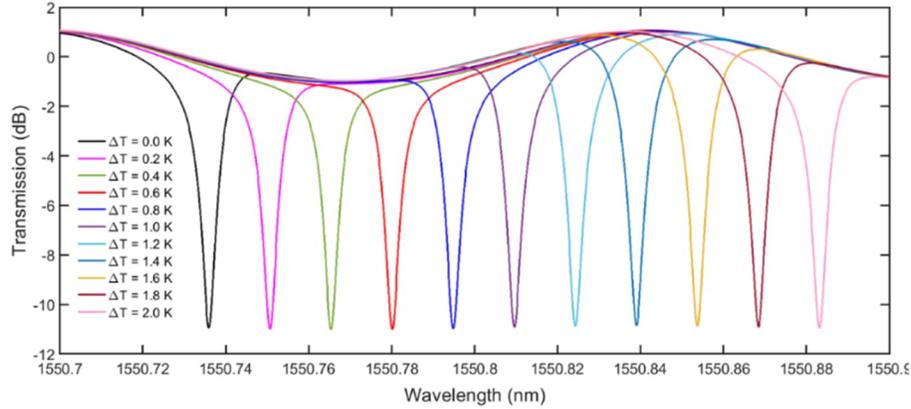

**Fig. 10.** The calculated transmission spectra for a silicon ring with length L=950 μm embedded in a 2 mm long FP cavity formed by two reflectors with R= 0.15, at several different ring temperature increments ΔT with ΔT= 0 corresponding to room temperature.

Fig. 10 shows the calculated transmission spectra for a ring with length L=950 μm, and is embedded in a 2 mm long FP cavity formed by two reflectors with reflectivity of R = 0.15, i.e., in the configuration shown in Fig. 7b. This model ring is nominally identical to ring B using the parameters in Table 1. The spectra are shown for several ring temperatures incremented in steps of 0.2 K. An asymmetrical skewing of the line shape is particularly evident when the resonance center coincides with the steepest slope of the FP fringes. Since the fluctuations vary with wavelength, the error in measured resonance wavelength centre is wavelength dependent, and therefore temperature dependent. The measured centre wavelength will periodically swing backwards (blue-shift) and forwards (red-shift) from the true resonance centre as the resonance moves relative to the back-ground ripple pattern. When used as a thermometer, the temperature readings will exhibit a corresponding periodic error as the temperature changes.

Fig. 11a shows an example simulation of the temperature reading error and resonance wavelength error for the ring-FP model used to create the spectra in Fig. 9. The resonance wavelengths plotted in Fig. 11 are extracted from the ring-FP spectra using three different measurement procedures. The first method determines resonance wavelength using a least-square error fit of a rigid line shape, with functional form as in Eq. 1, to the ring-FP spectrum at any given temperature, over a wavelength window spanning one -3 dB resonance width $\Delta\lambda_{3dB}$. The second procedure was the same except the fitting was carried out over a narrower one-half $\Delta\lambda_{3dB}$ width window, to reduce the range of the spectrum near the resonance minimum that contributes to the fit. Finally, the resonance positions were also determined simply by choosing the minimum point of resonance. This last approach gives a periodic error that is always within 0.02 pm of the resonance wavelength at each temperature, and hence the corresponding temperature error is less than 300 µK. The larger the fitting window over which the resonance is sampled, the stronger the effect of line distortions become. The 0.2 pm (3 mK) error obtained by fitting over the full -3 dB width window is much larger than using the smaller fitting window or simply taking the minimum transmission point. This behaviour arises because the ring DC transmission of Eq. 6 attenuates the intensity of the circulating FP reflections on each pass. This attenuation is highest near resonance, essentially turning off the FP cavity so that near the minimum point the resonance is only very weakly perturbed. The minimum point itself therefore gives the best estimate of the true resonance position. On the other hand, when fitting over a finite window, the extended line shape distortion contributes to the fit and creates a spurious measurement offset.

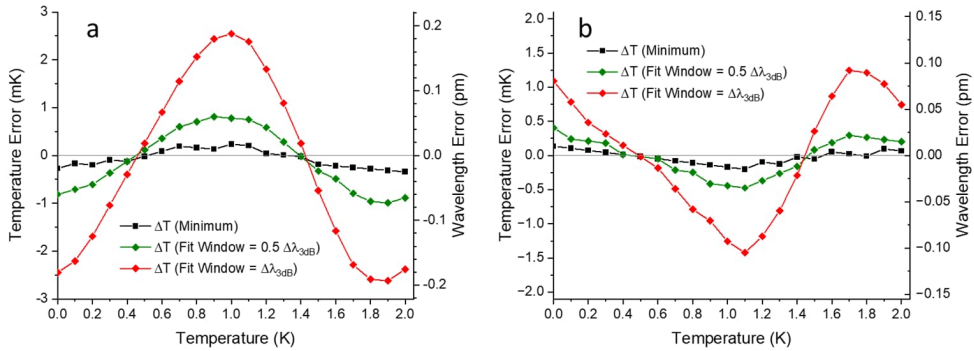

**Fig. 11.** (a) The error in resonance wavelength and temperature over a 2 K temperature range, for a ring of length L= 950 µm embedded in a 2 mm long FP cavity formed by two R=0.15 back-reflectors as in Fig. 8. The resonance wavelengths are determined by either fitting the resonance to an ideal line shape over windows with 3 dB line width $\Delta\lambda_{3dB}$, one-half $\Delta\lambda_{3dB}$, or simply by taking the minimum point of the resonance line shape. (b) Wavelength and temperature error as in (a) but with the ring located external to the FP cavity.

The temperature and resonance wavelength error for the back-reflection geometry of Fig. 9(c) is shown in Fig. 11(b). Here the ring is external to the FP cavity and does not itself modify the intensity fluctuations produced by the multiple back-reflections. The apparent resonance is skewed by the background intensity modulation, producing a resonance wavelength error that can be positive or negative depending on the relative alignment of the background modulation pattern and the resonance. The errors are smaller than in Fig. 11(a) because the ring and FP are not coupled and the phase distortion by the ring is not present. As in the coupled FP-ring geometry, the error is minimized by fitting to a narrow region around the resonance minimum or simple choosing the minimum point. The most accurate result for the scenarios of Fig. 9(b) and Fig. 9(c) would be expected for a critically coupled ring with a transmission zero on resonance, and simply tracking the wavelength of the zero-transmission point. Multiple internal reflections forming a parasitic FP cavity will be completely attenuated on resonance and the transmission zero will be the true resonance wavelength. The wavelength of a transmission zero also cannot be changed by any external intensity wavelength dependence or temporal modulation. These considerations suggest that second derivative based minimum detection methods combined with laser locking may useful in tracking thermometer resonances.

While the two configurations of Fig. 9(b) and 9(c) are very specific geometries, a complex spectral distortion can be considered a superposition of the back-reflection induced modulation patterns similar to these two examples. The results of Fig. 10 show that in either case the most accurate resonance measurement is obtained by measuring very near the resonance minimum, and preferably designing the ring to be critically coupled so that the resonance minimum is near zero. In practice, line shape fitting over a finite spectral window is often used to determine the center wavelength of a spectral line in order to reduce the uncertainty due to spectral noise fluctuations and detector noise [5]. The results here show that in the case of ring resonators in the all-pass configuration, this can be counter-productive. For example, the small resonance blue shift measured for rings C and D shown in Fig. 6 were only resolved by measuring the minimum point of each resonance. In our original attempt at data analysis using line shape fitting to determine resonance wavelengths, the wavelength scatter was several times larger than that seen in Fig. 6. The above discussion assumes that there are only intrinsic time independent signal distortions present. The measured resonance spectrum may also be changed by time varying external system noise from the photodetector, light source, and mechanical vibrations, and also by shot noise. Fitting may obviously improve measurement accuracy when such time varying external fluctuations are the dominant error source. In each measurement situation there will be a compromise to be considered before deciding on the optimal sensor data analysis strategy. Fitting may be helpful to mitigate time dependent noise, but should be carried out over a narrow spectral window when background spectral intensity variations and ripple are present. Averaging repeated spectral scans over the resonance position, combined with identifying the resonance minimum directly may be a best strategy to eliminate time dependent noise.

The numerical simulations used to generate the error curves in Fig. 11 assumed that the ring alone changed temperature while the through waveguide and grating couplers that contribute to background intensity fluctuations did not change temperature. In the case where the entire chip is at a uniform temperature, and all the waveguides have exactly the same thermo-optic coefficient, the ripple pattern in Fig. 10 will move rigidly in lockstep with the ring resonance as chip temperature changes. This is similar to the invariance of the back reflection induced line splitting observed in Fig. 8. The resonance wavelength displacement in this case will simply be a constant offset, and can be included in the calibration if it is possible to regard the entire chip as the effective thermometer element. This approach would not work when there are temperature dependent on-chip thermal gradients, or when the back reflection sources are not on the chip but elsewhere in the optical measurement path well away from the measurement point. Fig. 9(a) shows the resonance for ring C at T= 30 °C and T=80 °C, and it confirms that the grating coupler induced FP ripple does move almost in lockstep with the resonance line. There is a small 4 pm lag in the ring resonance shift relative to the ripple, which may be attributed to the small difference in the temperature dependent effective index of the curved ring waveguides and directional coupler, compared with the straight through waveguide that forms the FP cavity. Although this small lag would have an unmeasurable effect on the measured resonance minimum in these experiments, the would be compensated in the thermometer calibration procedure.

## 6. Summary

In this work we have examined some of the intrinsic sources of temperature measurement errors in silicon photonic ring resonator thermometers. These impairments may arise from the optical induced changes in the silicon waveguide, the primary example being optical absorption leading to local heating of the waveguide. Spectral distortions may also arise due to unintended reflections from defects, side-wall roughness, and discontinuous boundaries between different elements along the optical path such as from fiber facets and in-line polarizers or splitters, and also from grating couplers or waveguide facets at the chip edge.

Although the incident photon energy is well below the silicon band gap, refractive index changes may still arise from defect and surface state mediated absorption [26-30]. For ring cavities of 100 um or less, self-heating of the silicon waveguide is found to cause the ring waveguide to be several hundred mK hotter than the ambient environment even when using probe powers near 1 mW as is common in routine photonic device testing. At very low incident powers (< 20 μW) where self-heating is small, there remains a small blue shift of the ring resonance in all the Si devices tested, which corresponds to temperature reading errors near 2 mK. Such a small offset is not of concern in most applications but may be important for precision measurement and metrology standards work requiring sub-millikelvin resolution and accuracy.

The obvious mitigation strategy to reduce the power dependent effective index change is to minimize the optical probe power, but power can only be reduced to the level where signal-to-noise ratio becomes compromised by the detector noise and shot noise. An alternative method is to use a broad band light source and measure the ring spectrum using an external spectrometer. Since the power is distributed over a wide bandwidth, the circulating power at the resonance wavelength can be much lower than for a single line laser source, and self-heating will be correspondingly lower. Achieving comparable accuracy to our tunable lasers would require spectrometers with a resolution of better than 1 pm, which to best of our knowledge are not commercially available. The comparison of self-heating in short (Rings A and C) and long (Rings B and D) rings shows that simply increasing ring cavity length is a much simpler solution and does not require any change in measurement procedure. Measurements on Ring D show that the observed intensity dependent wavelength change can be less than less than 0.2 pm (i.e., < 3 mK temperature error) at probe powers that give more than adequate signal to noise. As previously noted, the presence of defect and surface states is likely the most significant cause of absorption related heating and index change in these waveguides. It is therefore expected that self-heating will be fabrication dependent, so the selection and optimization of fabrication processes to minimize near-surface defects and improve Si-cladding interface properties is also a path for temperature sensor improvement [28].

Back-reflections and back-scattering within the ring cause resonance splitting and line shape distortion. This is particularly common in high-Q, narrow linewidth resonators. Observed line shape distortions can vary randomly from one resonance order to the next in the same ring, but in this work we have found that the distorted line shapes are completely temperature independent due the identical temperature scaling of scattered field and primary ring mode propagation phases. Any such distorted resonance therefore remains suitable for tracking temperature as long as the fitting or peak finding method can reliably track resonance position. In the ring devices used in this work, severe line distortions and splitting are evident at some orders, but symmetric single lobed Lorentzian resonances are still common and such resonances remain unchanged with temperature and can serve as a reliable marker for temperature read out. Eliminating line splitting altogether requires reducing side wall roughness and designing the transitions between the DC, waveguide bends and straight waveguides to be sufficiently adiabatic that back reflections are negligible. Alternatively, the ring through waveguide coupling can be increased to increase the ring round trip loss relative to the back scattering, but this would sacrifice wavelength/temperature resolution since resonance line width increases with coupling strength.

The input and output surface grating couplers are particularly problematic since gratings designed for near normal coupling angle will have a grating periodicity near the Bragg condition for back-reflection, and therefore naturally form a weak FP cavity. Again, grating couplers can be designed with local structure factors that suppress such back-reflection [12,38]. Another simple design strategy to reduce ripple distortion is to shorten the distance between the input and output grating couplers. This increases the period of the FP ripple relative to the ring linewidth so the slope of background intensity is smaller with less distortion of the ring line shape. The improvement in background intensity modulation is illustrated by comparing the reflection induced ripple for a 2 mm coupler separation in Fig. 9(a) (Ring C), and that in Fig. 8 (Ring D) for which the coupler separation was reduced to approximately 900 μm.

Back-reflections and the resulting spectral distortion may also arise from off-chip optical elements, so similar care must be taken in building the measurement system and packaging the silicon photonic sensor chip. The sealed thermometer probe configuration described in Section 2 above, in which the fiber is several hundred micrometers away from the chip, is particularly effective in eliminating reflections from the input and output fiber facets and silicon surface. Since the free space beam is strongly diverging, these reflections are only very weakly coupled back into the fiber or the silicon chip with more than -15 dB attenuation and so will have little effect on the measured ring spectrum. In Fig. 9a the only visible spectral distortion is the fringe pattern of on-chip FP cavity formed by the input and output coupler, whereas measurements on similar devices in the direct fiber-coupled test bench configuration often exhibit more random background intensity variations as wavelength is scanned.

The resonance line shape very near the minimum wavelength point is relatively unperturbed by back-reflection induce spectral distortion, both when the ring is coupled to the reflection induced FP cavity and when there is a simple intensity modulation. In the case of a critically coupled ring the resonance minimum is zero and its wavelength is immune to spectral intensity distortions. Therefore, attempts at line shape fitting over a wide wavelength window may give poorer accuracy than simply determining resonance wavelength from the minimum point of the line. The experimental results in Fig. 4b demonstrate that accuracy down to 0.1 pm can be achieved by measuring the minimum point, even though the 3 dB width of the resonance is of the order of 10 pm.

In this work we have shown that silicon ring resonance wavelengths can be accurately measured and monitored down to the 0.2 pm level using commercial photonic test equipment and applying the mitigation strategies described above. In the application of such rings to photonic thermometers, this corresponds to a temperature accuracy of approximately 3 mK. In principle the mitigation methods suggested here may be extended to give a measurement resolution into the sub-millikelvin regime. Nevertheless, the role of near surface and interface states in optical absorption and carrier excitation implies that the precise nonlinear optical behaviour of each Si device may depend on details of fabrication history. Achieving accuracy and reproducibility to the 100 μK level or better will require individual device calibration, but this is also the current practice in case of SPRTs. Finally, it is important to consider the limits of state-of-the-art in photonic measurement and device fabrication. Obtaining ring resonance linewidths of less than a few picometers requires a level of control of the directional coupler dimensions and waveguide losses that is difficult to achieve with commonly used standard Si fabrication processes. For such very high Q resonators, line splitting and shape distortion will become more common as discussed in Section 6. Furthermore, the sub-picometer spectral resolution required to carry out the measurements in this paper is at the limits of commercial photonic test equipment (e.g., a tunable laser or spectrometer). Entering the microkelvin measurement regime using simple single ring devices will require spectral resolution, stability and accuracy in the femtometer (fm) wavelength range, and hence new and customized optical measurement systems. Alternatively, it may be possible to adopt more complex photonic chip architectures that amplify temperature response through, for example waveguide group index engineering [42]. Another approach that has only begun to be explored is to use the evolution of complex spectral patterns generated by two or more rings different thermo-optic coefficients [43, 44] to extract temperature.

**Disclosures.** The authors declare no conflicts of interest.

**Data availability.** Data underlying the results presented in this paper are not publicly available at this time but may be obtained from the authors upon reasonable request.

**References**

1. S. Dedyulin, Z. Ahmed, G. Machin, "Emerging technologies in the field of thermometry," Meas. Sci. Technol. **33**, 092001, 1-26 (2022).

2. N. Klimov, T.P. Purdy, Z. Ahmed, "Towards replacing resistance thermometry with photonic thermometry," Sens. Actuators A Phys. **269**, 308–312 (2018)
3. G. D. Kim, H. S. Lee, W. J. Kim, S. S. Lee, B. T. Lim, H. K. Bae, W. G. Lee, "Photonic temperature sensor based on an active silicon resonator manufactured using a CMOS process," Opt. Express **18**, 22215-22221 (2010).
4. S. Dedyulin, A. Todd, S. Janz, D.-X. Xu, S. Wang, M. Vachon and J. Weber, "Packaging and precision testing of fiber-Bragg-grating and silicon ring-resonator thermometers: current status and challenges," Meas. Sci. Technol. **31**, 074002, 1-7 (2020).
5. R. Eisermann, S. Krenek, G. Winzer, and S. Rudtsch, "Photonic contact thermometry using silicon ring resonators and tuneable laser-based spectroscopy," Technisches Messen **88**, 640–654 (2021).
6. J. Wang, Y. Pan, J. Gao, C. Zhang, Z. Qu, T. Xu, Y. Shen, J. Qu, "An on-chip silicon photonics thermometer with milli-Kelvin resolution," Appl. Sci. (MDPI) **12**, 3713, 1-8 (2022).
7. D.-X. Xu, A. Delâge, P. Verly, S. Janz, S. Wang, M. Vachon, P. Ma, J. Lapointe, D. Melati, P. Cheben and J.H. Schmid, "Empirical model for the temperature dependence of silicon refractive index from O to C band based on waveguide measurements," Opt. Express **27**(19), 27229-27242 (2019).
8. S. Dedyulin, A. Grzetic-Muffo, S. Janz, D.-X. Xu, A. D. W. Todd, M. Vachon, S. Wang, R. Cheriton, J. Weber, "Progress on silicon photonic thermometry for secondary and working measurement standards," in Photonics North Conference Proceedings, PN 2022, 9908405 (2022).
9. Consultative Committee for Thermometry under the auspices of the International Committee for Weights and Measures 2015, "Guide to the Realization of the ITS-90; Part 5—Platinum Resistance Thermometry," (Paris: BIPM), https://bipm.org/en/committees/cc/cct/guides-to-thermometry (2015).
10. H. Xu, M. Hafezi, J. Fan, J.M. Taylor, G.F. Strouse and Z. Ahmed, "Ultra-sensitive chip-based photonic temperature sensor using ring resonator structures," Opt. Express **22**, 3098-3104 (2014).
11. M. Novarese, S. Romero Garcia, S. Cucco, D. Adams, J. Bovington, and M. Gioannini, "Study of nonlinear effects and self-heating in a silicon microring resonator including a Shockley-Read-Hall model for carrier recombination," Opt. Express **30**, 14341-14357 (2022).
12. Y. Wang, X. Wang, J. Flueckiger, H. Yun, W. Shi, R. Bojko, N. A. F. Jaeger, and L. Chrostowski, "Focusing sub-wavelength grating couplers with low back reflections for rapid prototyping of silicon photonic circuits," Optics Ex. **22**(17), 20652 (2014)
13. "Birefringence control using stress engineering in silicon-on-insulator waveguides," W.N. Ye, D.-X. Xu, S. Janz, P. Cheben, M.-J. Picard, B. Lamontagne, N.G. Tarr, J. Lightwave Technol. **23**, 1308 (2005).
14. S. Dedyulin, A. Grzetic-Muffo, S. Janz, D.-X. Xu, S.Wang, M. Vachon and J. Weber, "Practical ring-resonator thermometer with an uncertainty of 10 mK," submitted (2023).
15. S.L. Gilbert and W.C. Swann, "Acetylene C2H2 Absorption Reference for 1510 nm to 1540 nm Wavelength Calibration—SRM 2517a," Natl. Inst. Stand. Technol. (US) Special Publication 260-133, ST (2001).
16. S. L. Gilbert, W. C. Swann, C.-M. Wang, Hydrogen Cyanide H13C14N 455 absorption reference for 1530nm to 1565nm wavelength calibration –456 SRM 2519a, Natl. Inst. Stand. Technol. (US) Special Publication 260-137, US 457 Government Printing Office: Washington, DC, (2005).
17. A. Yariv, "Universal relations for coupling of optical power between microresonators and dielectric waveguides" Electronics Lett. **36**(4), 321-322 (2000).
18. W. Bogaerts, P. De Heyn, T. Van Vaerenbergh, K. De Vos, S. Kumar Selvaraja, T. Claes, P. Dumon, P. Bienstman, V. Van Thourhout, and R. Baets. "Silicon microring resonators," Laser and Phot. Rev. **6,** 47-73 (2011)**.**
19. A.A. Nikitin, A.V. Kondrashov, V.V. Vitko, I.A. Ryabcev, G.A. Zaretskaya, N.A. Cheplagin, D.A. Konkin, A.A. Kokolov,L.I. Babak , A.B. Ustinov, B. A. Kalinikos, "Carrier-induced optical bistability in the silicon micro-ring resonators under continuous wave pumping," Opt. Commun. **480**, 126456, 1-6 (2021)
20. V. Jeyaselvan and S. Kumar Selvaraja, "Mitigation of carrier induced optical bistability in silicon ring resonators," Opt. Commun. **493**, 127021, 1-5 (2021)
21. R. Nandi, A. Goswami, and B. Krishna Das, "Phase Controlled Bistability in Silicon Microring Resonators for Nonlinear Photonics," IEEE J. Sel. Top. in Quantum Electron. **27**, 6100409 (2021).
22. W. Dickmann, L. Weituschat, R. Eisermann, S. Krenek, P.A. Postigo, S. Kroker, "Heat dynamics in optical ring resonators," in Proc. SPIE, Modelling Aspects in Optical Metrology VIII, 11783, 1178309 (2021).
23. L. Zhang, A.M. Agarwal, L.C. Kimerling and J. Michel, "Nonlinear Group IV photonics based on silicon and germanium: from near-infrared to mid-infrared," Nanophotonics, **3**(4-5), 247–268 (2014)
24. B. Kuyken, F. Leo, S. Clemmen, U. Dave, R. Van Laer, T. Ideguchi, H. Zhao, X. Liu, J. Safioui, S. Coen, S.P. Gorza, S. K. Selvaraja, S. Massar, R. M. Osgood Jr., P. Verheyen, J. Van Campenhout, R. Baets, W.M. J. Green, and G. Roelkens, "Nonlinear optical interactions in silicon waveguides," Nanophotonics," **6**(2), 377–392 (2017).
25. "Performance optimization of a reconfigurable waveguide digital optical switch on InGaAsP/InP: design, materials and carrier dynamics," S. Ng, S. Janz, W.R. McKinnon, P. Barrios, A. Delâge, and B.A. Syrett, IEEE J. Quantum Electron. **43** (12), 1147-1158 (2007).
26. D.E. Hagan and A.P. Knights, "Mechanisms for Optical Loss in SOI Waveguides for Mid-infrared Wavelengths around 2 mm," J. Opt. **19**, 025801 (2017)
27. M. Borselli, T.J. Johnson, and O. Painter, "Beyond the Rayleigh scattering limit in high-Q silicon microdisks: theory and experiment," Optics Exp. **13**, 1515-1530, (2005).

28. M. Borselli T.J. Johnson, and O. Painter "Measuring the role of surface chemistry in silicon microphotonics," Appl. Phys. Lett. **88**, 131114, 1-3 (2006).
29. T. Baehr-Jones, M. Hochberg, and A. Scherer, "Photodetection in silicon beyond the band edge with surface states," Opt. Exp.**16**, 1659-1668 (2008).
30. T. Baehr-Jones, M. Hochberg, and A. Scherer, "All-optical modulation in a Silicon Waveguide Based on a Singe Photon Process," IEEE J. Sect. Top. in Quant. Electron. **14**, 1335-1342 (2008).
31. Q. Li, A. Eftekhar, Z. Xia, A. Adibi, "Azimuthal Order Variations of Surface-roughness-induced mode splitting in high-Q Microdisk Resonators,' Opt. Lett. **37**(9) 1586-1588 (2012).
32. A. Li, T. Van Vaerenbergh, P. De Heyn, P. Bienstman, and W. Bogaerts, "Backscattering in silicon microring resonators: a quantitative analysis," Laser Photonics Rev. 10, No. 3, 420–431 (2016).
33. B.E. Little, J.-P. Laine, "Surface Roughness induced contradirectional coupling in ring and disk resonators," Opt. Lett. **22**(1) 4-6 (1997)
34. M.L. Gorodetsky, A.D. Pryamikov, V. S. Ilchenko, "Raleigh Scattering in High-Q Microspheres," J. Opt. Soc. Am. B **17**(6) 1051-1057 (2000).
35. T.J. Kippenberg, S.M. Splillen, and K.J. Vahal, "Modal Coupling in Travelling Wave Resonators," Opt. Lett. **27**(19), 1669-1671 (2002).
36. H. A. Haus and W. Huang, "Coupled-Mode Theory," Proc. of the IEEE **19**(10), 1501-1518 (1991).
37. C. Manolatou, M. J. Khan, Shanhui Fan, Pierre R. Villeneuve, H. A. Haus, , and J. D. Joannopoulos, "Coupling of Modes Analysis of Resonant Channel Add–Drop Filters," IEEE J. Quantum Electron. **35** (9), 1322-1331 (1999).
38. M. Kamandar Dezfouli, Y. Grinberg, D. Melati, P. Cheben, J.H. Schmid, A. Sánchez-Postigo, A. Ortega-Moñux, G.Wangüemert-Pérez, R. Cheriton, S. Janz, and D.-X. Xu, "Perfectly vertical surface grating couplers using subwavelength engineering for increased feature sizes," Opt. Lett. **45** (13), 3701-3704 (2020).
39. W. Liang, L. Yang, J.K.S. Poon, Y. Huang, K.J. Vahala, and A. Yariv, "Transmission characteristics of a Fabry–Perot etalon–microtoroid resonator coupled system," Opt. Lett. **31** (4), 510-512 (2006).
40. Z. Ahmed, N. Klimov, T.P. Purdy, T. Herman, K. Douglass, R. Fitzgerald, and R. Kundagrami, "Photonic Thermometry: Up-ending the 100 Year-Old Paradigm in Temperature Metrology," Proc. SPIE 10923, Silicon Photonics XIV, 109230L (2019).
41. G.R Fowles, "Introduction to Modern Optics," Dover Publications; 2nd ed. edition (June 1, 1989)
42. Y. Zhang, J. Zou and J.-J. He, "Temperature sensor with enhanced sensitivity based on silicon Mach-Zehnder interferometer with waveguide group index engineering," Optics Exp. **26**(20), 26057-26064 (2018).
43. S. Janz, R. Cheriton, D.-X. Xu, A. Densmore, S. Dedyulin, A. Todd, J.H. Schmid, P. Cheben, M. Vachon, M. Kamandar Dezfouli, and D. Melati, "Photonic temperature and wavelength metrology by spectral pattern recognition," Optics Exp. **28**(12), 17409-17423 (2020).
44. H.-T. Kim and M. Yu, "Cascaded ring resonator-based temperature sensor with simultaneously enhanced sensitivity and range," Optics Exp. **24**(9), 9501-9510 (2016).